\documentclass[epj]{svjour}
\usepackage{graphics}
\usepackage{epsfig}

\newcommand{\ds}{\displaystyle}
\newcommand{\dsf}{\ds\frac}
\newcommand{\Tr}{\mbox{Tr}}
\newcommand{\re}[1]{(\ref{#1})}
\newcommand{\no}{\nonumber}
\begin{document}
\title{Neutron-proton mass difference
in isospin asymmetric nuclear matter}

\author{
 Ulf-G. Mei{\ss}ner\inst{1,2}
   \and 
 A.~M.~Rakhimov\inst{3,4}
    \and 
 A.~Wirzba\inst{2}
    \and 
 U.~T.~Yakhshiev\inst{2,5}
}                     
%
%
\institute{
Helmholtz-Institut f{\" u}r Strahlen- und Kernphysik (Theorie),
D-53115, Universit{\" a}t Bonn, Germany
\and
Forschungszentrum
J{\" u}lich, Institut f{\" u}r Kernphysik
(Theorie),  D-52425  J{\" u}lich, Germany
\and
Institute of Nuclear Physics, Academy of Sciences of
Uzbekistan, Tashkent-132, Uzbekistan
\and
 Institute of Physics and Applied Physics,
Yonsei University, Seoul, 120-749, Korea
\and
Physics Department and Institute of Applied Physics,
National University of Uzbekistan, Tashkent-174, Uzbekistan}
\date{Received: date / Revised version: date}
%
\abstract{Isospin-breaking effects in the baryonic sector are studied in the
  framework of a medium-modified Skyrme model. The neutron-proton mass
  difference in infinite, asymmetric nuclear matter is discussed. In order to
  describe the influence of the nuclear environment on the skyrmions, we
  include energy-dependent charged and neutral pion optical potentials in the
  $s$- and $p$-wave channels.  The present approach predicts that the
  neutron-proton mass difference is mainly dictated by its strong part and
  that it strongly decreases in neutron matter.}



%
\maketitle
%

\section{Introduction}
\label{sect:intro}
\setcounter{equation}{0}

The medium dependence of isospin-breaking effects belongs to one of the
fundamental questions in nuclear 
physics \cite{Li:1997px,Baran:2004ih,Steiner:2004fi}.
In particular, the neutron-proton
mass difference 
in nuclear matter 
$\Delta m_{\rm np}^*$ is an interesting topic
of nuclear astrophysics relevant to the evolution of the universe at an early
stage~\cite{Steigman:2005uz,Cyburt:2004cq}.  
Furthermore,
it is also important for the description of the 
properties of mirror nuclei~\cite{Nolen:1969ms},
the stability of drip-line nuclei~\cite{Woods:1997cs}
and the transport in neutron-rich matter induced by 
heavy-ion collisions~\cite{Rizzo:2003if}.
Although there exist various publications dealing directly 
with the density dependence of the neutron-proton mass
difference
\cite{H:90pj,H:90zj,Dr:04,A:91js,M:91if,F:91qf,S:93xt,S:94tq,Chr:96}
and its implications for 
asymmetric nuclear
matter and finite nuclei 
properties
\cite{Bo91,Zu1,Zu5,Ch97,Le6,LQ88,vD5,vDC5,vD6,Ku97,Li01,Gr0,Ho0,Ts99,Ch7}, 
this quantity is still not well understood. Quantitatively and even
qualitatively
the predictions about the behavior of 
the neutron-proton mass difference in nuclear matter
change from model to model.  

Skyrme-soliton models have
the inherent advantage compared with other hadronic
models that they are based on chiral input and that they
treat the properties and interactions of the nucleons on an equal
footing~\cite{Skyrme:1961vq,Skyrme:1962vh}.
In this context we recently 
studied isospin-breaking effects in the baryonic sector of a
medium-modified Skyrme model~\cite{Meissner:2006id}, 
by focusing on the single hadron properties
in the nuclear environment 
rather than on the properties of the system as a whole.
The approach predicted that the neutron-proton mass difference 
changes in an isospin-symmetric nuclear environment by a very small amount.
The isospin-breaking leading to this result 
was only due to a modification of  
the mesonic sector
of the Skyrme model, originally introduced to generate 
the strong neutron-proton 
mass splitting in free space  
(in addition to the electromagnetic one)~\cite{Rathske:1988qt}.
However, when the nucleons are 
embedded in an isospin-asymmetric environment,
{\em additional} medium effects can be  expected.  To evaluate the latter 
in the present work we will consider the nucleon properties in  homogeneous, 
infinite and isospin-asymmetric nuclear matter. 
The Lagrangian of the present study is a generalization 
of the  in-medium Skyrme-type 
Lagrangian of Ref.~\cite{Meissner:2006id}. 
In addition to the strong isospin-breaking in the mesonic sector  it  
{\em explicitly}
takes into account the different influences of the 
isospin-asymmetric environment on the charged ($\pi^\pm$) and 
neutral ($\pi^0$) pion fields via a built-in energy dependence in the 
$s$-wave pion-nucleon ($\pi^\pm{\rm N}$) 
scattering 
lengths~\cite{Ber:93,Ber:95,Kaiser:2001bx,Kolomeitsev:2002gc,Friedman:2003ji}
and the $p$-wave 
optical potential~\cite{Ericsonbook}. This additional energy dependence
also alters the predictions for iso-sym\-metric matter considerably.

The paper is organized as follows:
In Sect.~\ref{sect:model} we formulate the 
model including the optical
potential input and medium modifications. Section~\ref{sect:classical}
discusses the classical Lagrangian and the pertinent equation
of motion. In Sect.~\ref{sect:mass_difference} we present the quantization
procedure and the final expressions for the the strong and electromagnetic
part of the in-medium neutron-proton mass difference. The results of the
calculation are reported and discussed in Sect.~\ref{sect:results}.
In Sect.~\ref{sect:summary} our conclusions are summarized and an outlook
to future studies is given. Moreover, for clarification, two appendices
are added. In the first one, we discuss in detail the setup of 
the classical equation of motion 
in the 
presence of isospin-breaking terms. 
The second appendix is devoted to peculiarities 
in
the construction of the charges and magnetic moments.

\section{Formulation of the problem}
\label{sect:model}

\subsection{Medium modification of the model}
\label{subsect:modification}

As in our previous work~\cite{Meissner:2006id}, we start with
a generalized Skyrme-type Lagrangian which incorporates
an explicit isospin-breaking term  in the mesonic 
sector:
\begin{eqnarray}
{\cal L}&=&{\cal L}_2+{\cal L}_4+{\cal L}_{{\rm g}\chi {\rm SB}}\,,
\label{lb}\\
{\cal L}_2&=&-\dsf{F_\pi^2}{16}\,\Tr\left( L_\mu L^\mu\right)\,,\\
{\cal L}_4&=&\dsf{1}{32e^2}\,\Tr\,[L_\mu,L_\nu]^2\,,\\
{\cal L}_{{\rm g}\chi {\rm SB}}&=&-\dsf{F_\pi^2}{16}\Bigl\{
 \Tr\left[(U-1){\cal M}_+^2(U^\dagger-1)\right]\quad\no\\
&&\quad\ \ \mbox{}-\Tr\left[(U-1)
\tau_3{\cal M}_-^2 (U^\dagger-1)\tau_3\right]\Bigr\},
\label{LRathske}
\end{eqnarray}
where Einstein's summation convention is always assumed 
(if not specified otherwise).
$L_\mu=U^\dagger \partial_\mu U$ is given in terms of  the chiral 
$SU(2)$ matrix
$U=\exp(2i \tau_a \pi_a/F_\pi)$, where $\pi_a$  ($a=1,2,3$) are the 
Cartesian isospin-components of the pion field.
$F_\pi=2f_\pi$ is the pion-decay constant, while $e$ is the dimensionless
Skyrme constant. Finally, 
${\cal M}_\pm$ $\equiv$ $\sqrt{(m_{\pi^\pm}^2\pm m_{\pi^0}^2)/2}$ is defined 
in terms of the
masses of the charged and neutral pions.
As in Ref.~\cite{Meissner:2006id} we insist 
on reproducing the empiri\-cal (isospin-averaged)
masses of the nucleon and delta,
$m_{\rm N} = 938$~MeV and $M_{\Delta} = 1232$~MeV, in free space 
(density $\rho=0$)
and without isospin breaking term (${\cal M}_-=0$).
Furthermore, as input for the free mass of the neutral pion
we take the PDG-value~\cite{Yao:2006px}:
$m_{\pi^0}=134.977$~MeV. These choices 
induce the values $F_{\pi}= 108.11$~MeV  and $e = 4.826$. 
Following Ref.~\cite{Meissner:2006id}
the mass 
of the charged pions  $m_{\pi^\pm}$
is extracted as a variational parameter  $m_{\pi^\pm}=135.015$~MeV
from the fit to the empirical value $\Delta m_{\rm np}^{(exp)}=1.29$~MeV
of the neutron-proton mass splitting in free space. Note that 
the dominant electromagnetic contribution to  $m_{\pi^\pm}-m_{\pi^0}$
is beyond the scope of the model.

The generalized pion mass term~\re{LRathske}, which was originally proposed 
by Rathske~\cite{Rathske:1988qt}, can be rewritten as
\begin{eqnarray}
{\cal L}_{{\rm g}\chi {\rm SB}}&=&-\dsf{F_\pi^2}{16}
\bigg\{\Tr\left[(U-1)m_{\pi^0}^2(U^\dagger-1)\right]\no\\
&&\quad\ \ 
\mbox{}+\sum_{a=1}^2\Tr(\tau_aU)\,{\cal M}_-^2\,\Tr(\tau_aU^\dagger)\bigg\}\,,
\label{gchiSB}
\end{eqnarray}
which is convenient for our up-coming modifications.
When the pion fields are small, 
\begin{equation}
U=\exp\left\{\dsf{2i\vec\tau\cdot\vec \pi}{F_\pi}\right\}
\approx
1+\dsf{2i\vec\tau\cdot\vec\pi}{F_\pi}
+\dots\,,
\label{linapp}
\end{equation}
the Lagrangian~\re{lb} reduces to the Lagrangian for free pion fields
\begin{eqnarray}
{\cal L}_{\rm low}&=& 
\partial_\mu\pi^+\partial^\mu\pi^--\pi^+m_{\pi^\pm}^2\pi^-\no\\
&&\mbox{}+\dsf{1}{2}
\left(\partial_\mu\pi^0
\partial^\mu\pi^0-\pi^0 m_{\pi^0}^2\pi^0\right)\,,
\label{llowfree}\\
\pi^\pm&=&\dsf{1}{\sqrt2}(\pi_1\mp i\pi_2)\,,\quad
\pi^0=\pi_3\,.\end{eqnarray}
In the medium the analog  of the Lagrangian~\re{llowfree} reads
\begin{eqnarray}
{\cal L}_{\rm low}^*&=&
\dsf{1}{2}\sum_{\lambda=\pm,0}
\big\{\partial_\mu{\pi^\lambda}^\dagger
\partial^\mu\pi^{\lambda}-{\pi^{\lambda}}^\dagger\big(m_{\pi^\lambda}^2+
\hat\Pi^{\lambda}\big)\pi^\lambda\big\}\no\quad\\
&=&{\cal L}_{\rm low}-\dsf{1}{2}\Big\{\pi_a\hat\Pi^{0}\pi_a
+i\varepsilon_{ab3}\pi_a\Delta\hat\Pi\pi_b\Big\}\,, 
\label{llowmed}
\end{eqnarray}
where 
$\hat\Pi^0$ and $\Delta\hat\Pi$ are linear combinations of the self energies
of the charged pions:
\begin{eqnarray}
\hat\Pi^0&=&\dsf{1}{2}\big(\hat\Pi^{-}+\hat\Pi^{+}\big)\,,\no\\
\Delta\hat\Pi&=&\dsf{1}{2}\big(\hat\Pi^{-}-\hat\Pi^{+}\big)\,.
\label{Pi0DPi}
\end{eqnarray}
As in Ref.~\cite{Meissner:2006id} 
the medium-modified version (always mar\-ked  by an asterix) 
of the Lagrangian~\re{lb}, 
can be defined 
as
\begin{equation}
{\cal L}\to{\cal L}^*
={\cal L}_2+{\cal L}_4+{\cal L}_{{\rm g}\chi{\rm SB}}^*\,,
\label{lmed}
\end{equation}
where 
for the general case of asymmetric matter 
${\cal L}_{{\rm g}\chi{\rm SB}}^*$ is given by 
formula\footnote{{}From now on $m_\pi$ stands for the mass of the
neutral pion, {\it i.e.} $m_\pi\equiv m_{\pi^0}$.}
\begin{eqnarray}
{\cal L}_{{\rm g}\chi{\rm SB}}^*
&=&-\dsf{F_\pi^2}{16}\Bigl\{\Tr\Bigl[(U-1)
\big(m_{\pi}^2+\hat\Pi^{0}\big)(U^\dagger-1)\Bigr]\nonumber\\
&&\mbox{}+\sum_{a,b=1}^2
\Tr(\tau_aU)\,\Bigl[\delta_{ab}{\cal M}_-^2\no\\
&&\mbox{}\qquad+i\varepsilon_{ab3}
\Delta\hat\Pi/2\Bigr]\,\Tr(\tau_bU^\dagger)\Bigr\}\,.
\label{gchiSBmed}
\end{eqnarray}
It is easy to check that 
the  Lagrangian~\re{lmed} reduces to the Lagrangian~\re{llowmed} 
for the expansion~\re{linapp} as well as 
to the medium-modified
Lagrangian of Ref.~\cite{Meissner:2006id}
for the case of isospin-symmetric matter, $\hat\Pi^+=\hat\Pi^-=\hat\Pi^0$
(in the parameterization of Ref.~\cite{Meissner:2006id}).

\subsection{Parameterization of the optical potentials}

The polarization operators of the charged pions can be expressed 
in terms of energy-dependent 
pion-nucleus optical potentials~\cite{Ericsonbook} as follows:
\begin{eqnarray}
\hat\Pi^{\pm}_{s}(\omega,\vec r)&=&-4\pi
b^\pm(\omega,\vec r)\equiv \chi^{\pm}_{s}(\omega,\vec r)\,,
\label{chis}\\
\hat\Pi^{\pm}_{p}(\omega,\vec r)&=&
\vec\nabla
\dsf{4\pi c^\pm(\omega,\vec r)}{1\!+\!4\pi g^\prime c^\pm(\omega,\vec r)}
\cdot\vec\nabla
-\dsf{4\pi\omega}{2m_N}\left(\vec\nabla^2c^\pm(\omega,\vec r)\right)\no\\
&\equiv&\vec\nabla\chi^{\pm}_{p,1}(\omega,\vec r)\cdot\vec\nabla
-\dsf{\omega}{m_\pi}\chi^{\pm}_{p,2}(\omega,\vec r)\,,
\label{chip}\\
b^\pm(\omega,\vec r)
&\equiv&\Big(b_0^{\rm eff}(\omega)\rho(\vec r)\mp b_1(\omega)
\delta\rho(\vec r)\Big)\eta\,,\\
c^\pm(\omega,\vec r)&\equiv&\Big(c_0(\omega)\rho(\vec r)\mp 
c_1(\omega)\delta\rho(\vec r)\Big)\eta^{-1},\\
\rho(\vec r)&=&\rho_n(\vec r)+\rho_p(\vec r)\,,\\
\delta\rho(\vec r)&=&\rho_n(\vec r)-\rho_p(\vec r)\,,\\
\eta&=&1+m_{\pi}/m_{\rm N}\,,
\end{eqnarray}
where $\rho_n$ and $\rho_p$ are the neutron and proton densities, 
respectively.
Note that 
additional $\vec\nabla^2\rho$ and
 $\vec\nabla^2\delta\rho$ terms are included 
in the $p$-wave optical potential since they are 
needed for the description of  realistic pion-nucleus scattering 
data~\cite{Ericsonbook}.

The chiral expansion of the off-shell pion-nucleon scattering amplitudes 
at vanishing pion three-momentum leads to 
energy-dependent $s$-wave isoscalar 
and isovector scattering lengths,  $b_0(\omega)$ and $b_1(\omega)$, 
respectively.
The quantities 
$c_0(\omega)$ and $c_1(\omega)$ are the corresponding $p$-wave
scattering volumes, whereas $b_0^{\rm eff}(\omega)$ is the {\em effective}
isoscalar scattering length (see Eq.~\re{b0eff}).
The correlation
parameter $g'$, which renormalizes the pion dipole susceptibility, is fixed 
at $g^\prime=0.47$.

Within the counting scheme of pion-nucleon chiral perturbation theory from 
Refs.~\cite{Ber:93,Ber:95} and based on input from 
these references
$b_0(\omega)$ and $b_1(\omega)$ can be expressed
at  order ${\mathcal O}(m_\pi^3)$ 
as~\cite{Kaiser:2001bx,Kolomeitsev:2002gc}
\begin{eqnarray}
b_0(\omega)&\approx&\dsf{1}{4\pi\eta}\left(\dsf{\sigma_{\pi N}-\beta\omega^2}
{f_{\pi,\rm ph}^2}
+\dsf{3g_A^2m_{\pi}^3}{16\pi f_{\pi,\rm ph}^4}\right)
\no\\
&\approx&
 \dsf{1.206m_{\pi}^{-1}}{4\pi\eta}\big(1-m_{\pi}^{-2}\omega^2\big)\no\\
&\equiv& -\dsf{\tilde b_0}{4\pi\eta}\big(1-m_{\pi}^{-2}\omega^2\big)\,,
\label{b0omega}
\\
b_1(\omega)&\approx&-\dsf{1}{4\pi\eta}\,\dsf{\omega}{2f_{\pi,ph}^2}
\left(1+\dsf{\gamma\omega^2}{4\pi^2f_{\pi,\rm ph}^2}\right)
\no\\
&\approx&
 -\dsf{1.115m_{\pi}^{-1}}{4\pi\eta}
\big(m_{\pi}^{-1}\omega+0.143m_{\pi}^{-3}\omega^3\big)\no\\
&\equiv&\dsf{\tilde b_1}{4\pi\eta}
\big(m_{\pi}^{-1}\omega+0.143m_{\pi}^{-3}\omega^3\big)\,.
\label{b1omega}
\end{eqnarray}
Here $\sigma_{\pi N}=-4C_1m_{\pi}^2-9g_A^2m_{\pi}^3/64\pi f_{\pi,\rm ph}^2\approx
45$~MeV is the 
pion-nucleon
sigma term, whereas the other parameters correspond to  
the ``range term'' \cite{Delorme:1992cn,Thorsson:1995rj,Meissner:2001gz}
$\beta=g_A^2/4m_N-2C_2-2C_3\approx0.541m_{\pi}^{-1}$ 
and also $\gamma=(g_A\pi
f_{\pi,\rm ph}/m_{\rm N})^2 + \ln(2\Lambda_{\rm c}/m_{\pi}) \approx2.523$.
The axial-vector coupling constant $g_A=1.27$ and the pion decay constant
$f_{\pi,\rm ph}=92.4$~MeV are fixed to their empirical values, since
they refer to the parameterization of 
the nuclear (matter) background. On the other hand,
the parameter  
$F_\pi=2f_\pi$ of the Skyrme Lagrangian is fixed to the value 108.11~MeV,
since this parameter 
together with $e = 4.826$ and
$m_{\pi}=134.977$~MeV refers to the soliton itself; {\it i.e.} 
the empiri\-cal
(iso\-spin-ave\-raged) masses of the nucleon and delta are reproduced 
by this choice,  as explained 
below Eq.~\re{LRathske}.
The value of the cutoff-scale parameter $\Lambda_{\rm c}=737$~MeV is
adjusted to the the threshold value of the isospin-odd on-shell $\pi{\rm N}$ 
scattering amplitude. 
The dimension-two 
low-energy constants $C_{1,2,3}$ can be found {\it e.g.} in 
Refs.~\cite{Ber:93,Ber:95}. These values are consistent with
the most recent
 pion-nuclei scattering data as it was
summarized in Ref.~\cite{Meissner:2005ba} and  
lead to the threshold values $b_0(m_{\pi})\approx 0$ and
$b_1(m_{\pi})\approx-0.0883m_{\pi}^{-1}$, respectively~\cite{Schroder:2001rc}.
Within the errors  
these values of the scattering lengths are consistent with
the more refined analysis of
Ref.~\cite{Meissner:2005ne}.
Furthermore, the incorporation of 
double scattering corrections in the $s$-wave pion polarization
operator leads to the effective isoscalar scattering length
\begin{equation}
b_0^{\rm eff}(\omega)
\approx b_0(\omega)-\dsf{3k_F}{2\pi}\big[b_0^2(\omega)+2b_1^2(\omega)\big]\,,
\label{b0eff}
\end{equation}
where $k_F=[3\pi^2\rho/2]^{1/3}$ is the total Fermi momentum. 
The terms of higher order than $\omega^2$ 
can be neglected in $b_0^{\rm eff}$ and $b_1$, 
provided that the condition 
$\omega < m_{\pi}$ is met.\footnote{Within the framework of the Skyrme model,
this situation corresponds to nucleons with 
$S=T=\frac{1}{2}\sim \omega\Lambda$,
where $\Lambda\approx 1$\,fm  is the moment of inertia of the skyrmion. 
In the case of $\Delta$-isobar
states ($S=T=\frac{3}{2}\sim \omega\Lambda$) also 
$\omega^n$ terms with $n\ge 3$ have to be taken into 
account.\label{foot-omega}}

For simplicity, we ignore the energy dependence in the $p$-wave scattering
volumes and replace $c_0(\omega)$ and $c_1(\omega)$ 
by the constant threshold values 
$c_0(m_\pi)$ = $0.21m_\pi^{-3}$ and $c_1(m_\pi)$ = $0.165m_\pi^{-3}$ 
of the `current' SAID analysis~\cite{SAID}.
This is compatible with  the discussion in
Ref.~\cite{Friedman:2003ji}.
Furthermore,
all terms proportional to odd $\omega$ powers in
$\hat\Pi^0_{p}$ and even ones 
in $\Delta\hat\Pi_{p}$ are neglected. 
This is consistent with the remark in footnote~\ref{foot-omega} and
the disregard of
pion-absorption in this approach.

In summary,
one can write the polarization 
opera\-tors \re{Pi0DPi} as follows:
\begin{eqnarray}
\hat\Pi^0_{s}&=&\dsf{\chi_{s}^-(\omega)+\chi_{s}^+(\omega)}{2}
\no\\
&\approx&
\left(\tilde b_{0}+\dsf{3k_F}{8\pi^2\eta}\tilde b_{0}^2\right)\rho
-\left(\tilde b_0+\dsf{3k_F}{4\pi^2\eta}\left(\tilde b_0^2-\tilde b_1^2\right)
\right)\rho\dsf{\omega^2}{m_\pi^2}\no\\
&\equiv&\chi_{s}^{00}-\chi_{s}^{02}{m_\pi^{-2}}{\omega^2}\,,
\label{polOpb}
\\
\hat\Pi^0_{p}&=&
\vec\nabla\dsf{\chi_{p,1}^-(m_\pi)+\chi_{p,1}^+(m_\pi)}{2}\cdot\vec\nabla
\no\\ &\equiv&
\vec\nabla\chi_{p}^0\cdot\vec\nabla
\approx
\vec\nabla\frac{4\pi c_0(m_\pi)\rho/\eta}
{1+4\pi g'c_0(m_\pi)\rho/\eta} 
\cdot\vec\nabla,
\label{polOpb2}\\
\Delta\hat\Pi_s&=&- \tilde b_{1}\delta\rho\,{m_\pi^{-1}}\omega
\equiv -\Delta\chi_{s}{m_\pi^{-1}}\omega\,,
\label{polOpbe}\\
\Delta\hat\Pi_p&=&-\dsf{2\pi\omega}{m_N\eta}\, c_1(m_\pi)
\left(\vec\nabla^2\delta\rho\right)
\equiv -\Delta\chi_{p}{m_\pi^{-1}}\omega\,.
\label{polOpe}
\end{eqnarray}

\subsection{Medium-modified Lagrangian}

Evidently the explicit expressions  of the self energies in 
configuration space follow from the standard rules
$\hat\Pi^0(\omega)$ $\to$ $\hat\Pi^0(i\partial_0)$ and
$\Delta\hat\Pi(\omega)$ $\to$ $\Delta\hat\Pi(i\partial_0)$. 
After inserting the polarization operators 
$\Pi(i\partial_0)$ 
and  $\Delta\Pi(i\partial_0)$ from \re{polOpb}-\re{polOpe}
into the Lagrangian~\re{lmed} and integrating 
by part, in order to symmetrize in the time derivatives, one arrives at 
the final form of the medium-modified Lagrangian:
\begin{eqnarray}
{\cal L}^*&=& {\cal L}^*_{\rm sym}+{\cal L}^*_{\rm as}\,,
\label{lmedbeg}\\
{\cal L}^*_{\rm sym}&=&{\cal L}_2^*+{\cal L}_4+{\cal L}_{\chi{\rm SB}}^*\,,\\
{\cal L}^*_{\rm as} &=&
\Delta{\cal L}_{\rm mes}+\Delta{\cal L}_{\rm env}^*\,,\\
{\cal L}_2^* &=& \dsf{F_\pi^2}{16}
\Big\{\big(1+m_\pi^{-2}\chi_{s}^{02}\big)
\Tr\left(\partial_0U\partial_0U^\dagger\right)\no\\
&&\qquad\mbox{}-\left(1-\chi_{p}^0\right)
\Tr(\vec{\nabla} U\cdot\vec{\nabla} U^\dagger)
 \Big\}\,,\\
{\cal L}_{\chi {\rm SB}}^*&=&-\dsf{F_\pi^2 m_{\pi}^2}{16}
\big(1+m_\pi^{-2}{\chi_{s}^{00}}\big)
\no\\&&\mbox{}\times 
\Tr\left[(U-1)(U^\dagger-1)\right]\,,\\
\Delta{\cal L}_{\rm mes}&=&-\dsf{F_\pi^2}{16}\sum_{a=1}^2{\cal M}_-^2
\Tr(\tau_aU)\Tr(\tau_aU^\dagger),\qquad\\
\Delta{\cal L}_{\rm env}^*&=&-\dsf{F_\pi^2}{16}\sum_{a,b=1}^2
{\varepsilon_{ab3}(2m_\pi)^{-1}\left(\Delta\chi_{s}+\Delta\chi_{p}\right)}\no\\
&&\qquad\qquad\mbox{}\times \Tr(\tau_a U)\Tr(\tau_b\partial_0 U^\dagger)\,.
\label{lmedend}
\end{eqnarray}
Here $\Delta{\cal L}_{\rm mes}$  and  $\Delta{\cal L}^*_{\rm env}$
are the isospin-breaking terms 
arising from the 
explicit symmetry breaking in the 
mesonic sector and  the isospin asymmetry of the 
surrounding environment, respectively.

Note that both the temporal part of ${\cal L}_2^*$ and the chiral symmetry
breaking term ${\cal L}_{\chi {\rm SB}}^*$, decrease -- at leading order
linearly -- with increasing matter 
density, since $\chi_{s}^{02}$ and $\chi_{s}^{00}$ are negative, see 
Eqs.~\re{b0omega} and \re{polOpb}.
However, as the same equations indicate,  
$\chi_{s}^{02}\approx \chi_{s}^{00}\approx \tilde b_0\rho$, such
that the effective mass, determined by the mass pole of the in-medium  
propagator, is approximately 
density-independent in agreement with the findings about
the in-medium Gell-Mann--Oakes--Renner relation of 
Refs.~\cite{Thorsson:1995rj,Meissner:2001gz}.
Furthermore, one can 
see that the Lagrangian~\re{lmedend} contains the Weinberg-Tomozawa term,
as the relation 
\begin{equation}
\dsf{\Delta\chi_s}{4\pi\eta}=-\dsf{m_\pi\delta\rho}{8\pi\eta f_{\pi,\rm ph}^2}
=b_1^{\rm l.o.}\delta\rho
\label{WeinTomo}
\end{equation} 
is based on the isovector $s$-wave scattering length in the chiral 
expansion to lowest  order~\cite{Weinberg:1966kf,Tomozawa:1966jm}.

The Lagrangian~\re{lmedbeg}-\re{lmedend} will be used in our studies of
isospin breaking effects in asymmetric nuclear matter.
In the next sections we will present and discuss the changes 
that emerge due to the isospin asymmetry of the surrounding
nuclear environment.  Specifically, we will concentrate on
isospin-breaking effects in infinite nuclear matter 
with a constant density, so that  the
$p$-wave contribution proportional to  $\Delta\chi_p\sim
\vec\nabla^2\delta\rho$ vanishes. Note that in the case of finite
nuclei this term may be essential for  nucleons located near the surface
of the nucleus.

\section{Classical solitonic solutions}
\label{sect:classical}
By following the two-stage method of Ref.~\cite{Meissner:2006id}\footnote{
An alternative, but equivalent way of introducing this method   
is presented in Appendix~\ref{app:ClassSol}.}  
of constrained and unconstrained collective isospin-rotations,
applied to
the hedgehog ansatz 
$U=\exp\left[i\tau\cdot (\vec{r}/r) F(r)\right]$, 
the time-dependent Lagrangian can be constructed from Eq.~\re{lmedbeg} 
in terms of the standard angular velocities $\omega_i$ of the collective 
modes and the constrained angular velocity $a^*$ (see below) as
\begin{eqnarray}
L^*&=&\int {\cal L}^* {\rm d}^3 {r}
=-M_{\rm NP}^*-{\cal M}_-^2\Lambda_- +
\dsf{\vec\omega^2}{2}\Lambda^*\no\\
&&\mbox{}+\omega_3\big(a^*\Lambda^*+\Delta^*\big)+
a^*\left(\dsf{a^*}{2}\Lambda^*+\Delta^*\right).
\label{lag1}
\end{eqnarray}
Here
\begin{eqnarray}
M_{\rm NP}^*&=&\pi\int\limits_0^\infty\bigg\{
\dsf{F_\pi^2}{2}\left(1-\chi_{p}^0\right)\left(F_r^2 +\frac{2\,S^2}{r^2}\right)
\nonumber\\
&&\quad\mbox{}+
\dsf{2}{e^2}\left(2F_r^2+\frac{S^2}{r^2}\right)\frac{S^2}{r^2}
\nonumber\\
&&\quad\mbox{}+
F_\pi^2\left(m_\pi^2+{\chi_{s}^{00}}\right)
\left(1-\cos F\right)\bigg\}r^2\,{\rm d}r\qquad
\label{clEm0}
\end{eqnarray}
is the in-medium mass of the 
soliton when it is not perturbed (NP) by any isospin breaking. 
The abbreviations $F_r\equiv {\rm d} F/{\rm d} r$ and 
$S\equiv\sin F$ have been used, where $F=F(r)$ is the chiral profile function
of the hedgehog ansatz. 
Furthermore
\begin{equation}
\Lambda^*=\big(1+m_{\pi}^{-2}\chi_s^{02}\big)\Lambda_-+\Lambda_4
\label{momin}
\end{equation}
with the separate contributions
\begin{eqnarray}
\Lambda_-&=&\dsf{2\pi}{3}F_\pi^2\int\limits_0^\infty S^2\,
   r^2\,{\rm d}r\,,
\label{clEm}\\
\Lambda_4&=&\dsf{8\pi}{3e^2}\int\limits_0^\infty 
\left(F_r^2+\frac{S^2}{r^2}\right)S^2 \,r^2\,{\rm d}r
\end{eqnarray}
is  the in-medium moment-of-inertia, whereas 
\begin{equation}
\Delta^*=(2m_\pi)^{-1}\Delta\chi_s\Lambda_-
\label{clEe}
\end{equation}
is
the response of the isospin-asymmet\-ric environment (see Eqs.~\re{polOpbe}
and \re{lmedend}) to
the collective iso-rotations.

The constrained angular velocity parameter $a^*$ corresponds to a 
stationary rotation around the third axis in isotopic space that
serves to undo the effect of the mesonic isospin-breaking term
proportional to ${\cal M}_-$
at the classical level, when
the collective rotational modes in the isospin-space 
are frozen ($\omega_{1,2,3}\rightarrow 0$).
In this classical limit, 
applying the constraint~\cite{Meissner:2006id} 
\begin{equation}
a^{*2}=2{\cal M}_-^2{\Lambda_-}/{\Lambda^*}\,,
\label{constr}
\end{equation}
one generates the Lagrangian
\begin{equation}
L^*=-M_{\rm NP}^*+a^*\Delta^*\,.
\label{lclassic}
\end{equation}
The pertinent equation of motion for the hedgehog profile 
function $F(r)$
takes then the form
\begin{eqnarray}
&&F_\pi^2(1-\chi_p^0)
\left(r^2F_{rr}+{2}{r}F_r-{S_2}\right)\no\\
&&\mbox{}+\dsf{4}{e^2}\left[{2S^2}F_{rr}+{S_2}
\left(F_r^2-\dsf{S^2}{r^2}\right)\right]\no\\
&&\mbox{}-{F_\pi^2}\left(m_{\pi}^2+\chi_s^{00}\right)Sr^2
+a^*\dsf{F_\pi^2\Delta\chi_s}{3m_\pi}S_2r^2
=0\,,
\label{classic-eq}
\end{eqnarray}
where the additional abbreviations 
$S_2$ = $\sin 2F$ and $F_{rr}$ = $d^2F/dr^2$ 
were introduced.

The solution corresponding to the soliton of
baryon number $B=1$ fulfills the boundary conditions
\begin{eqnarray}
\lim_{r\rightarrow 0}F(r)&=&\pi-C r\,,\\
\lim_{r\rightarrow \infty}F(r)&=&{D}\left(1+m_\beta r\right)
\exp\left\{-m_\beta r\right\}/r^2\,,
\label{infasymp}\\
m_\beta^2&=&\dsf{m_\pi^2+\chi_s^{00}-
2a^*m_\pi^{-1}\Delta\chi_s/3}{1-\chi_p^0}\,,\qquad
\label{beta}
\end{eqnarray}
where $C$ and $D$ are constants.
Since the parameter $a^*$ is part of  the classical equation \re{classic-eq},
{\it i.e.} $h(F_{rr},F_r,F,a^*)=0$, 
this equation together with the constraint \re{constr}   can 
be solved by iteration:~\footnote{The choice of the 
sign of $a^*$ is fixed by  the sign
of $\Delta m_{\rm np}^{\rm strong}$ in free space~\cite{Meissner:2006id}.} 
\[
\begin{array}{lcl}
h\big(F_{rr}^{(0)},F_r^{(0)},F^{(0)},0\big)= 0&
\  \Rightarrow\  & a^*_0=a^*\big(F_r^{(0)},F^{(0)}\big)\,;\\
h\big(F_{rr}^{(n)},F_r^{(n)},F^{(n)},a_{n-1}^*\big)=0
&\  \Rightarrow\  & a^*_{n}=a^*\big(F_r^{(n)},F^{(n)}\big)\,.
\end{array}
\]
In the actual calculation, this iteration scheme rapidly converges
after 3 to 4 iteration steps.

\section{In-medium neu{\-}tron-pro{\-}ton mass difference}
\label{sect:mass_difference}

\subsection{Strong part of  $\Delta m_{\rm np}^*$}

By taking into account the condition~\re{constr},
applying the definition $a_{\rm eff}^*\equiv
a^*+{\Delta^{*}}/{\Lambda^{*}}$  and 
using the canonical quantization procedure as in Ref.~\cite{Meissner:2006id},
one can construct from the Lagrangian \re{lag1}
the Hamiltonian in terms of the isospin operator $\hat{\vec T}$:
\begin{eqnarray}
{\hat H^*}  &=&M_{\rm NP}^*
+\dsf{\hat{T_1}^2}{2\Lambda^*} 
+\dsf{\hat{T_2}^2}{2\Lambda^*}
+\dsf{\left(\hat{T_3}-\Lambda^* a_{\rm eff}^*\right)^2}{2\Lambda^*}\no\\
&=&M_{\rm NP}^*+\dsf{\hat{\vec{T}}^2}{2\Lambda^*}-
a_{\rm eff}^*\hat T_3+\Lambda^* \dsf{(a_{\rm eff}^*)^2}{2}\,.
\label{Ham}
\end{eqnarray}
Thus the strong part of the neutron-proton mass difference can be identified 
as 
\begin{equation}
\Delta m_{\rm np}^{*(\rm strong)}=a_{\rm eff}^*=
a^*+\dsf{\Delta^{*}}{\Lambda^{*}}\,.
\label{dm-strong}
\end{equation}
Note that the density-variation of the strong part 
of the neutron-proton mass difference
will be more pronounced than in Ref.~\cite{Meissner:2006id} for the following
reasons: (i) the 
explicit density-dependence of the moment 
of inertia $\Lambda^*$ (see Eq.~\re{momin}) resulting from the
energy-dependent
parameterization of the optical potentials, and (ii) the existence of the
additional term $\Delta^*/\Lambda^*$
in an isospin-asymmetric environment. Even if the explicit isospin 
breaking in the mesonic sector were omitted, ${\cal M}_-=0$, there still 
would be a non-vanishing neutron-proton 
mass splitting proportional to the isospin-asymmetric 
environment factor $\Delta^*$.

\subsection{Electromagnetic  part of  $\Delta m_{\rm np}^*$}

As discussed in  Appendix~\ref{app:Charges}, 
by calculating the pertinent 
Noether currents one can construct the following
isoscalar (S) and isovector (V) electromagnetic (EM) form factors
\begin{eqnarray}
G_{\rm E}^{{\rm S}*}({\vec q}^2)&=&\int\limits_0^\infty
\left(\dsf{\tilde B}{2}-\dsf{\Delta^*}{\Lambda^*}\,\tilde\Lambda+
\tilde\Delta^*\right) j_0(qr)\,{\rm d}r\,,
\label{ges}\\
G_{\rm M}^{{\rm S}*}({\vec q}^2)&=&\dsf{m_{\rm N}(1+\Delta^*)}{2\Lambda^*}
\int\limits_0^\infty\tilde Br^2\,\dsf{j_1(qr)}{qr}\,{\rm d}r\,,
 \label{gems}\\
G_{\rm E}^{{\rm V}*}({\vec q}^2)&=&\dsf{1}{2\Lambda^*} \int\limits_0^\infty
\tilde\Lambda^*j_0(qr)\,{\rm d}r\,,
\label{gev}\\
G_{\rm M}^{{\rm V}*}({\vec q}^2)&=&m_{\rm N}\int\limits_0^\infty
\left[\big(1-\chi_{\rm p}^0\big)\tilde\Lambda_-+\tilde\Lambda_4^*
\right]\dsf{j_1(qr)}{qr}\,{\rm d}r\,,\qquad
\label{gmv}
\end{eqnarray}
in terms of the spherical Bessel functions $j_0$ and $j_1$ and the
three-momentum transfer $q=|\vec q|$.
Here a quantity with a tilde, say   $\tilde Z=\tilde Z(F)$, is defined
as the integrand of the corresponding functional, {\it i.e.}:
\[
 Z[F]\equiv \int\limits_0^\infty \tilde Z\bigl(F(r)\bigr)\, {\rm d}r\,.
\] 
As usual,  $B = 1$ is 
the baryon charge, such that $\tilde B(r)$ = $4\pi r^2 B^0(r)$, where 
$B^0(r)$ = $-\sin^2 F F_r/(2\pi^2 r^2) $ is
the baryon density of the skyrmion. 
The medium-dependent form factors of the proton and neutron are defined
as 
\[
 G_{\rm E,M}^{\left({\rm p}\atop {\rm n}\right)*}({\vec q}^2)
=G_{\rm E,M}^{{\rm S}*}({\vec q}^2)
\pm G_{E,M}^{{\rm V}*}({\vec q}^2)
\]
with the normalization conditions
$G_{\rm E}^{{\rm p}*}(0)$ = $1$, $G_{\rm E}^{{\rm n}*}(0)$ = $0$, 
$G_{\rm M}^{{\rm p}*}(0)$ = $\mu_{\rm p}^*$, 
$G_{\rm M}^{{\rm n}*}(0)$ = $\mu_{\rm n}^*$,
where $\mu_{\rm p}^*$ and $\mu_{\rm n}^*$ are the 
magnetic moments of the in-medium proton and neutron,
respectively.

In the present approach all  form factors  {\em explicitly} depend on 
medium functionals, on the one hand, 
via the density-dependent moment of inertia
$\Lambda^*$ (see Eq.~\re{momin}),
and on the other hand, via additional terms resulting from the 
isospin-asymmetric nuclear environment. Moreover, note the additional terms
in the isoscalar form factors as compared with 
Ref.~\cite{Meissner:2006id}, which 
emerge here from that part 
of the isospin charge density that is independent of
the isospin $T_3$ (see Appendix~\ref{app:Charges}).

Finally, applying the formula~\cite{Gasser:1982ap}
\begin{eqnarray}
\Delta m_{\rm np}^{*(\rm EM)}&=&-\dsf{4\alpha}{\pi}\int\limits_0^\infty 
{\rm d}q
\bigg\{G_{\rm E}^{{\rm S}*}(\vec{q}^2)G_{\rm E}^{{\rm V}*}(\vec{q}^2)
\nonumber\\
&&\qquad\mbox{}-\dsf{\vec{q}^2}{2m_{\rm N}^2}
G_{\rm M}^{{\rm S}*}(\vec{q}^2)G_{\rm M}^{{\rm V}*}(\vec{q}^2)\bigg\}\,,
\qquad
\label{dm-EM}
\end{eqnarray}
where $\alpha\approx 1/137$ is the fine-structure constant,
one can calculate the medium-dependent 
electromagnetic part of the neutron-proton mass difference as in 
Ref.~\cite{Meissner:2006id}.

\section{Results and discussions}
\label{sect:results}

\begin{figure}[hbt]
\epsfysize=5.8cm
\epsffile{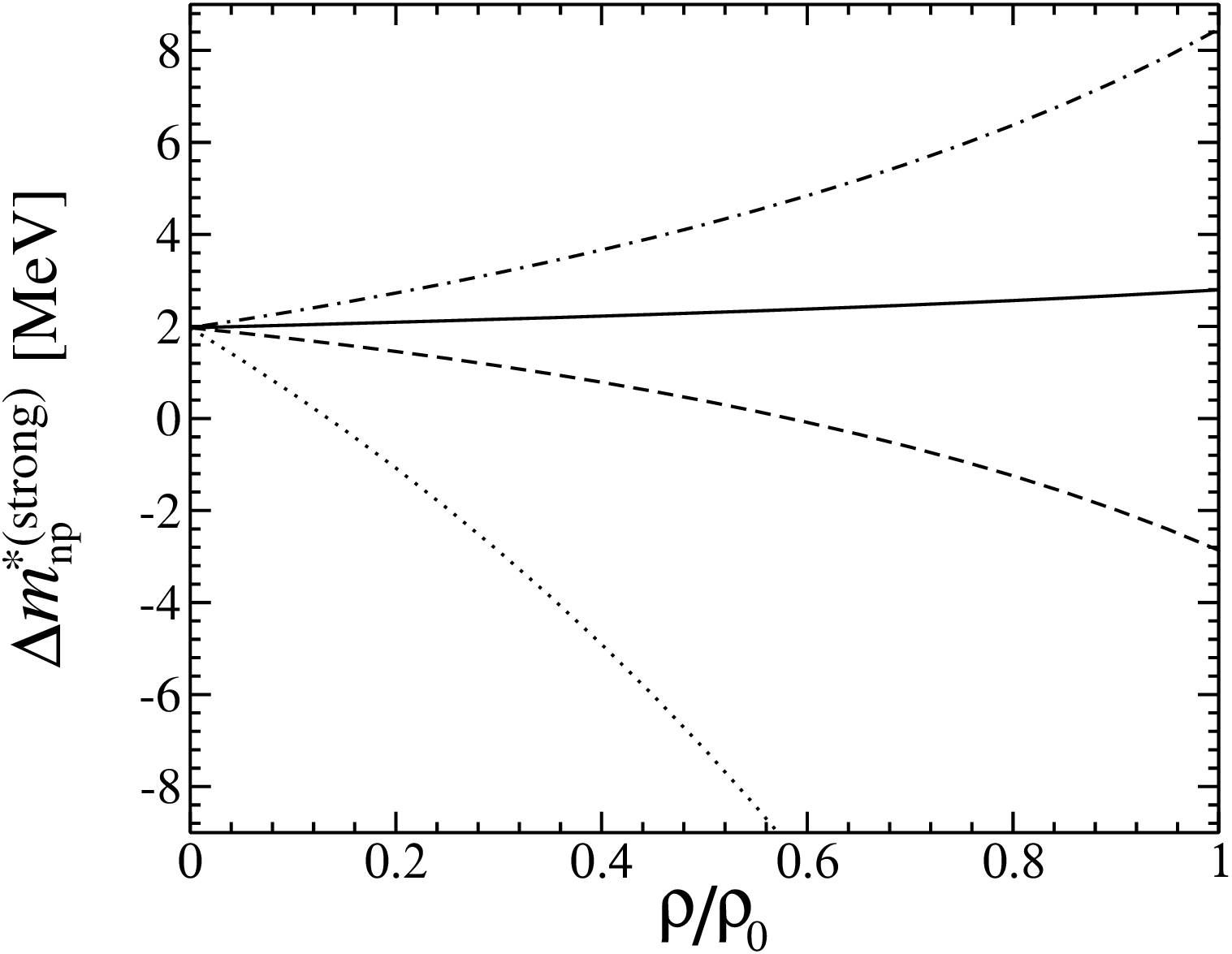}
\vspace{0.2cm}
\caption{Density dependence of the strong part $\Delta m_{\rm np}^{*(\rm
    strong)}$ of the neutron-proton mass difference.  The abscissa represents
  the density $\rho$ normalized to the saturation density of ordinary nuclear
  matter $\rho_0=0.5m_{\pi}^3$, while the ordinate shows the mass difference
  in units of MeV.  The result in isospin-symmetric matter is plotted as a
  solid curve, the result of neutron-rich matter with $\delta\rho/\rho=0.2$ 
  as dashed curve, the dotted curve represents pure neutron matter
  ($\delta\rho/\rho=1$) and the dot-dashed curve proton-rich matter with
  $\delta\rho/\rho=-0.2$.}
\label{fig-strong}
\end{figure}

In Fig.~\ref{fig-strong} the strong part of the in-medium 
neutron-proton mass splitting, $\Delta m_{\rm np}^*$, is shown
for isospin-symmetric nuclear matter
(solid curve), neutron-rich matter (dashed curve), pure neutron matter
(dotted curve), and
proton-rich matter (dot-dashed curve).
Already in iso\-spin-symme\-tric 
matter $\Delta m_{\rm np}^{*(\rm strong)}$ has visibly 
a different density-behavior than the corresponding quantity of 
Ref.~\cite{Meissner:2006id}.
For example, at normal nuclear matter density,
$\Delta m_{\rm np}^{*(\rm strong)}$ has increased 
by about 42\% relative to its 
free space value (see the solid curve of Fig.~\ref{fig-strong}). 
In contrast to this in the previous 
work~\cite{Meissner:2006id}, 
where the optical potentials were assumed to be energy-independent,
$\Delta m_{\rm np}^{*(\rm strong)}$ decreased by a very tiny amount, namely by
about $2\%$ at normal nuclear matter densities; in other words,
$\Delta m_{\rm np}^{*(\rm strong)}$ in Ref.~\cite{Meissner:2006id} 
was practically density-independent.

Moreover, when the isospin symmetry of nuclear matter is broken,
 $\delta\rho/\rho\ne0$,\footnote{This quantity may be called 
the isospin-asymmetry parameter of the nuclear environment.}
$\Delta m_{\rm np}^{*(\rm strong)}$ strongly varies  at the qualitative as well 
as quantitative level  (see the dashed curve in Fig.~\ref{fig-strong}).  
In pure neutron matter the change becomes very drastic 
(see the dotted curve in Fig.~\ref{fig-strong}), and 
$\Delta m_{\rm np}^{*(\rm strong)}$ 
decreases very fast with increasing density. 

\begin{figure}
\epsfysize=5.8cm
\epsffile{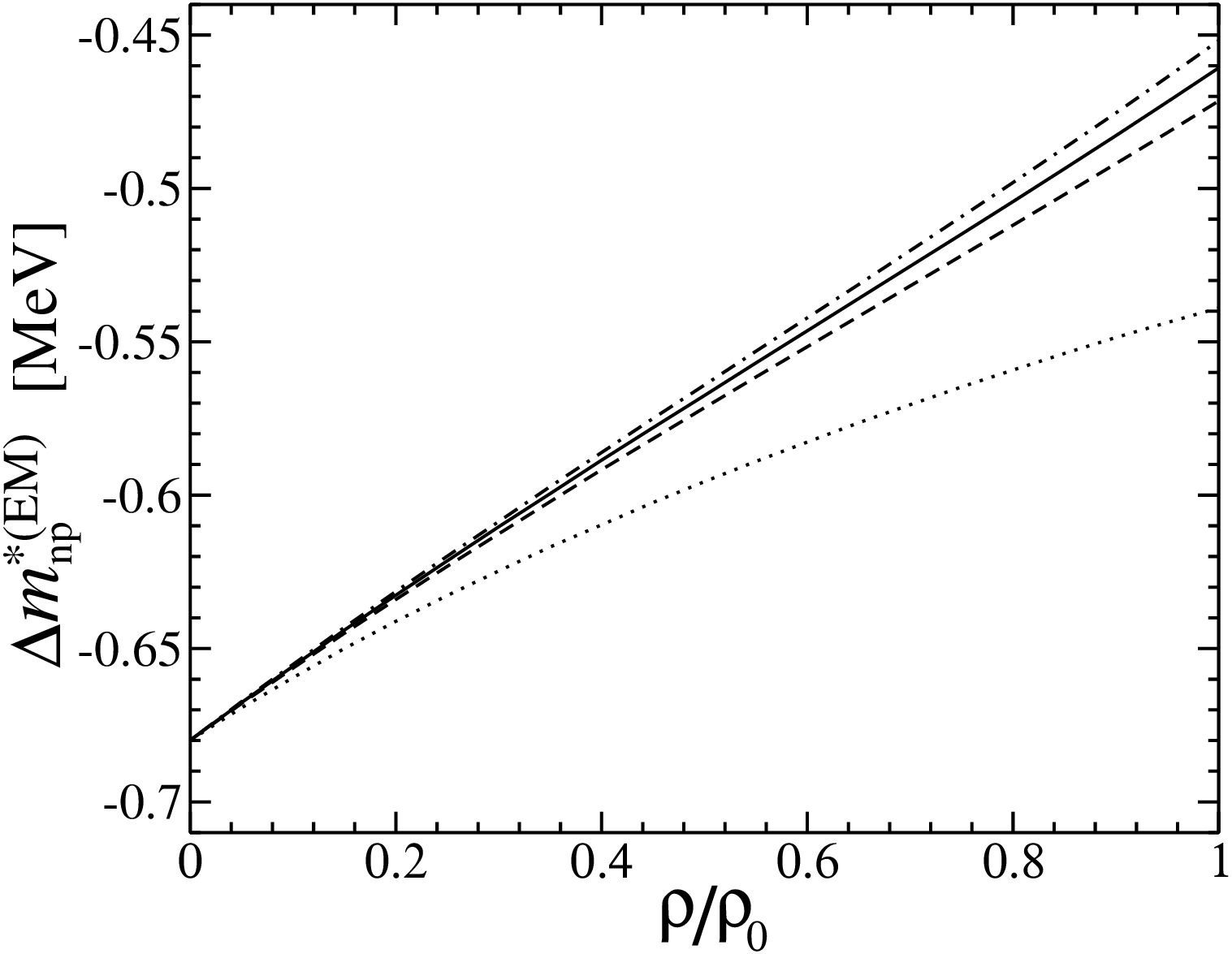}
\vspace{0.2cm}      
\caption{Density dependence of the electromagnetic part $\Delta m_{\rm
    np}^{*(\rm EM)}$ of the neutron-proton mass difference. The axes and
  curves are defined as in Fig.~\ref{fig-strong}.}
\label{fig-EM}
\end{figure}

In  contrast to the strong part, 
the electromagnetic part of the neutron-proton mass 
difference varies only by a small amount when the isospin-asymmetry parameter 
is increased (see Fig.~\ref{fig-EM}).
But compared with the result of 
the previous approach~\cite{Meissner:2006id}, the change
is sizable, even in isospin-symmetric matter. This is again
due to the explicit density
dependence of the moment of inertia~\re{momin}, 
and the changes in the 
solutions of the
classical equations~\re{classic-eq}.
Note that with increasing 
density the moment of inertia $\Lambda^*$ 
decreases since 
$\chi_s^{02}<0$. In addition, the solutions of the classical
equations~\re{classic-eq} are altered because  $\chi_s>0$ in Eq.~(22) of
Ref.~\cite{Meissner:2006id} is replaced by the $\omega$-independent
part of the present Eq.~\re{polOpb}, 
namely by $\chi_s^{00}$ which is negative.

\begin{figure}
\epsfysize=5.8cm
\epsffile{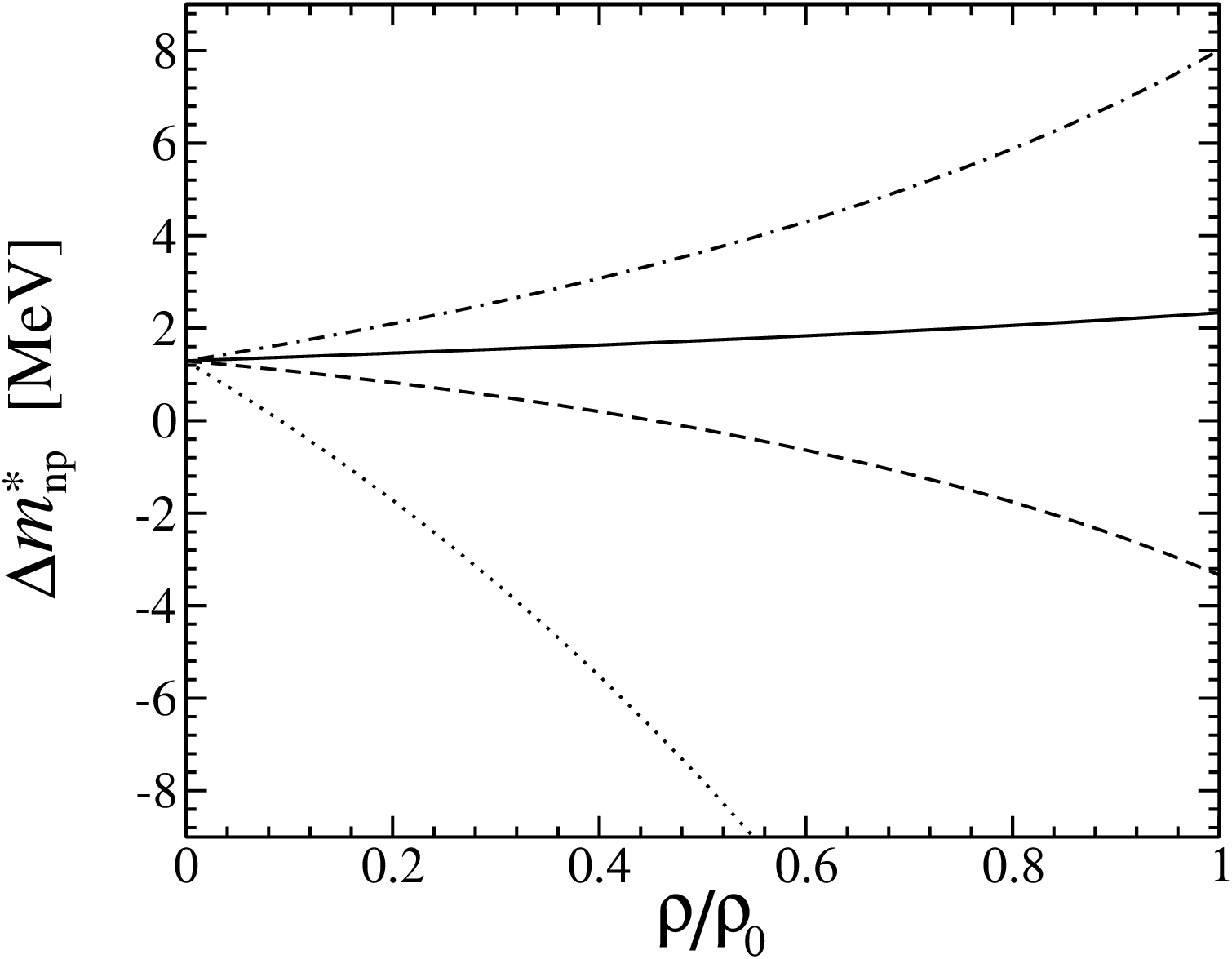}
\vspace{0.2cm}      
\caption{
   Density dependence of the total neutron-proton mass difference
  $\Delta m_{\rm np}^{*}$.  The axes and curves are defined as in
  Fig.~\ref{fig-strong}.}
\label{fig-total}
\end{figure}

For completeness, we present the total neutron-proton 
mass difference in Fig.~\ref{fig-total}.
{}From a comparison with Fig.~\ref{fig-strong} it is obvious that
this mass  difference is completely dominated by its strong
part.
In pure neutron matter and at the density $\rho_0$,
the neutron-proton mass difference is $\Delta m_{\rm np}^*
=-25$~MeV. For comparison, the authors of the work~\cite{Dr:04} 
got the result $\Delta m_{\rm np}^*\approx-70$~MeV
in framework of QCD sum rule studies.

Another interesting result is the difference between 
the values of $\Delta m_{\rm np}^{*}$ in neutron-rich and proton-rich 
matter -- compare the dashed and dash-dotted curves in Fig.~\ref{fig-total}.
One can see that in neutron-rich matter 
$\Delta m_{\rm np}^{*}$ is decreased relative to 
the isospin-symmetric case, whereas in proton-rich matter 
the behavior is just opposite.
This finding may become useful for future studies of 
mirror nuclei and their properties. 
For example, for the case of the mirror nuclei $^{48}\rm Ca$ and $^{48}\rm Ni$,
a similar behavior of $\Delta m_{\rm np}^*$ 
was found in Ref.~\cite{Ho0}  
within density-dependent relativistic hadron field theory.
The result of our work is also consistent with the findings  of 
Refs.~\cite{LQ88,vD5,vDC5,vD6,Ku97,Li01,Gr0} 
that utilize a 
relativistic approach  and with
the nonrelativistic calculation~\cite{Ch97} 
based on Skyrme-like  effective interactions.

Even at the qualitative level, the 
various models mentioned in the introduction 
differ in their predictions of the 
neutron-proton mass difference
in nuclear matter: 
(i) in nonrelativistic approaches~\cite{Bo91,Zu1,Zu5},
which are focused on the system properties as a whole,
this difference mainly turns out to be positive  ($\Delta m^*_{\rm np}>0$); 
(ii) however, it is negative ($\Delta m^*_{\rm np}<0$) 
in relativistic approaches
\cite{LQ88,vD5,vDC5,vD6,Ku97,Li01,Gr0}
and some nonrelativistic 
variational calculations~\cite{Ch97} or it becomes negative
with increasing density in QCD sum rule 
studies \cite{H:90pj,H:90zj,Dr:04};
(iii) it depends 
on the isospin content of the system ($\Delta m^*_{\rm np} > 0$ or 
$\Delta m^*_{\rm np} < 0$)
in relativistic hadron field theory~\cite{Ho0}. 
The effective masses 
in relativistic approaches are discussed in detail 
in Ref.~\cite{Jaminon:1989wj}. Furthermore,  
the difference in the behavior of $\Delta m_{\rm np}^*$ 
in the relativistic and nonrelativistic approaches is explained
in Ref.~\cite{vD5}. 

\begin{figure}
\epsfysize=5.8cm
\epsffile{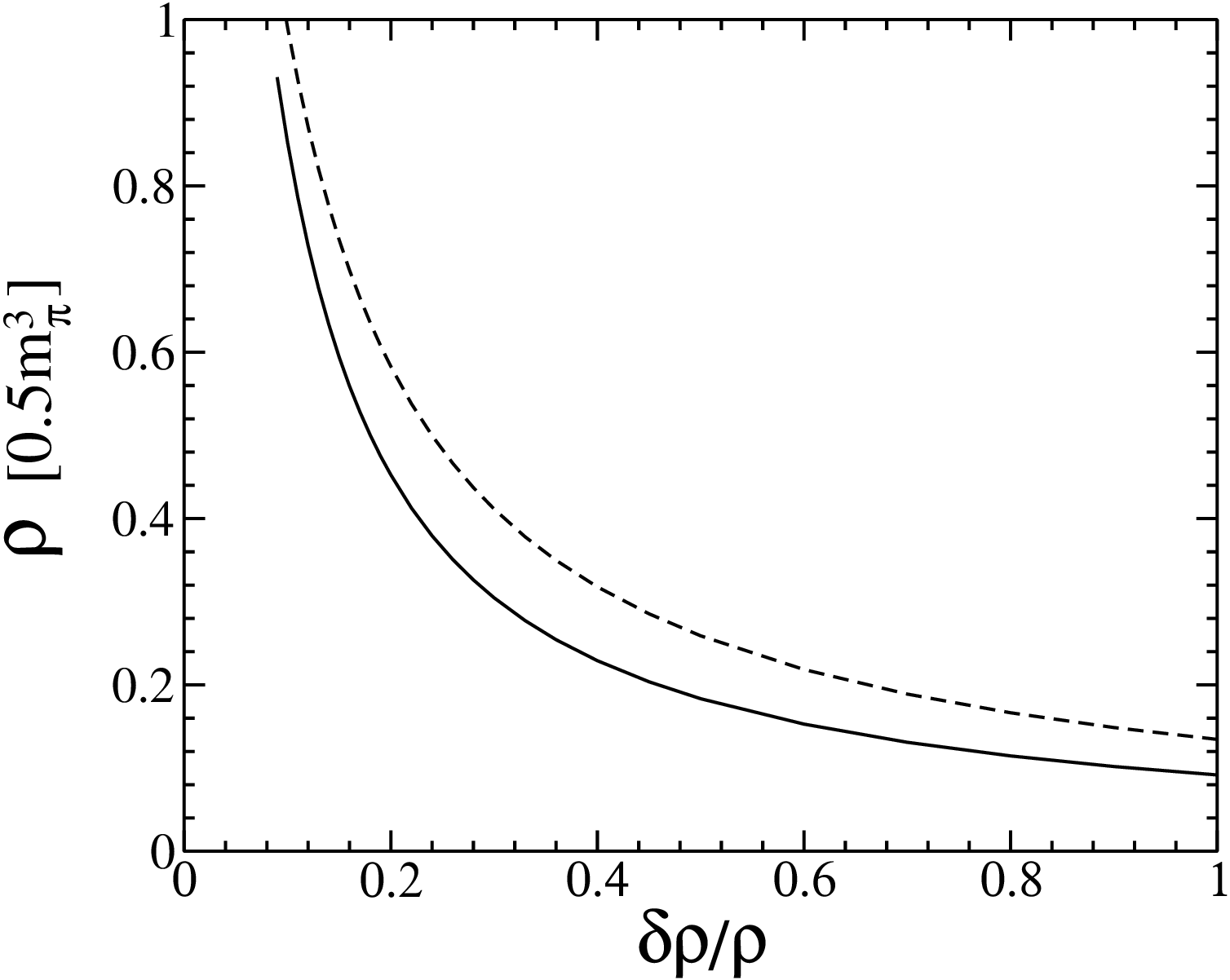}
\vspace{0.2cm}      
\caption{The solutions of the 2-parameter equations $\Delta m_{\rm
    np}^*(\rho,\delta\rho)=0$ and $\Delta m_{\rm
    np}^{*\rm(strong)}(\rho,\delta\rho)=0$. The abscissa represents the
  isospin-asymmetry parameter, while the ordinate shows the density (in units
  of the ordinary nuclear matter density $\rho_0=0.5m_{\pi}^3$), 
  where the neutron-proton
  mass difference (solid curve) or its strong part (dashed curve) vanishes.}
\label{fig-as-rho}
\end{figure}

Also our approach shows  that $\Delta m^*_{\rm np}$ qualitatively 
depends on the isospin content of surrounding environment. It
is always positive
in  proton-rich matter as well as  in isospin-symmetric matter
(see the solid and dot-dashed curves in Fig.~\ref{fig-total}). 
In neutron-rich matter, however, the sign may  change.
For the reader's convenience, we plot
in Fig.~\ref{fig-as-rho}  
those values of the density $\rho$ as function of the
isospin-asymmetry parameter $\delta\rho/\rho$ where the in-medium
neutron-proton mass splitting or its strong part 
vanishes. In other words the solutions 
of the 2-parameter 
equations, $\Delta m_{\rm np}^*(\rho,\delta\rho)=0$
and $\Delta m_{\rm np}^{*\rm(strong)}(\rho,\delta\rho)=0$, are presented.
For small positive input for the isospin-asymmetry parameter the
neutron-proton mass difference or its strong part 
vanishes at high densities (of the order
of the ordinary nuclear matter density $\rho_0$).
With increasing $\delta\rho/\rho$, however, the mass difference changes 
its sign at moderate densities, and in strongly isospin-asymmetric
matter  this transition is already at low densities.
For instance,
in neutron-rich matter with the isospin-asymmetry parameter
$\delta\rho/\rho\sim 0.1$ the 
proton becomes heavier at the density $\rho\sim 0.85\rho_0$.
In pure neutron matter this change happens 
already at the density  $\rho\sim 0.09\rho_0$.

In addition, in Table~\ref{table1} we present 
the calculated effective masses and
isoscalar as well as isovector charge radii of the in-medium nucleons 
for some values of the nuclear matter density.
\footnote{Note that the
tabulated values of the free proton and neutron mass differ from their 
PDG values~\cite{Yao:2006px} since the  
customary Skyrme value 
$M_N=938\,{\rm MeV}$ was used here and in Ref.~\cite{Meissner:2006id}
as input for the isospin-averaged nucleon mass.}
\begin{table}
  \caption{Calculated masses (in units of MeV) and 
    isoscalar as well as isovector charge radii 
    (in units of fm) of the nucleons in 
    nuclear matter of density $\rho$ (in units of
    the saturation density of  ordinary nuclear matter 
    $\rho_0=0.5m_{\pi}^3$).
  } \label{table1}
\begin{center}
\begin{tabular}{ccccc}
\noalign{\smallskip}
\hline
\noalign{\smallskip}
$\rho/\rho_0$&$m_{\rm p}^*$\,&$m_{\rm n}^*$\,&
$\langle r^2\rangle^{*1/2}_{\rm E,I=0}$\,&
$\langle r^2\rangle^{*1/2}_{\rm E,I=1}$\\
\noalign{\smallskip}\hline\noalign{\smallskip}
\multicolumn{5}{c}{In free space}\\
0      &937.4\,\,&938.7\,\,& 0.49 & 0.74\\
\noalign{\smallskip}\hline\noalign{\smallskip}
\multicolumn{5}{c}{In proton-rich matter ($\delta\rho/\rho=-0.2$)}\\
0.5\,\,\,&729.6\,\,\,&733.3\,\,& 0.61 & 0.84\\
1.0\,\,\,&547.9\,\,\,&555.9\,\,& 0.79 & 0.98\\
\noalign{\smallskip}\hline\noalign{\smallskip}
\multicolumn{5}{c}{In isospin symmetric matter ($\delta\rho/\rho=0$)}\\
0.5\,\,\,&729.7\,\,\,&731.4\,\,& 0.60 & 0.84\\
1.0\,\,\,&547.9\,\,\,&550.2\,\,& 0.75 & 0.98\\
\noalign{\smallskip}\hline\noalign{\smallskip}
\multicolumn{5}{c}{In neutron-rich matter ($\delta\rho/\rho=0.2$)}\\
0.5\,\,\,&731.5\,\,\,&731.3\,\,& 0.58 & 0.84\\
1.0\,\,\,&553.4\,\,\,&550.0\,\,& 0.72 & 0.98\\
\noalign{\smallskip}\hline\noalign{\smallskip}
\multicolumn{5}{c}{In pure neutron matter ($\delta\rho/\rho=1$)}\\
0.5\,\,\,&757.9\,\,\,&750.1\,\,& 0.54 & 0.83\\
1.0\,\,\,&632.1\,\,\,&607.0\,\,& 0.52 & 0.94\\
\noalign{\smallskip}\hline
\end{tabular}
\end{center}
\end{table}
In general, the nucleon masses strongly  decrease in the nuclear medium
and are qualitatively in agreement with the well known 
results~\cite{Jeukenne:1976uy,Mahaux85}. 
At normal nuclear matter density and for an isospin asymmetry
$\delta\rho/\rho\sim0.25$,
the difference in the  effective masses (normalized to  the 
corresponding free space values) of the neutron and
proton,
respectively, 
is  $m^*_{\rm n}/m_{\rm n}-m^*_{\rm p}/m_{\rm p}\sim 0.01$. 
For comparison, the result of Ref.~\cite{Ho0} 
for nucleons located near the center of $^{132}\rm Sn$
is one order of magnitude bigger:
$m^*_{\rm n}/m_{\rm n}-m^*_{\rm p}/m_{\rm p}\sim 0.1$.

The isoscalar and isovector charge radii 
$\langle r^2\rangle^{*1/2}_{\rm E,I=0,1}$
increase with increasing 
density of the medium.\footnote{Note that our results in free space 
differ from the ones of Ref.~\cite{Adkins:1983hy} by a factor 
$\sqrt{2}$ due to 
different normalizations of the charge densities (see Eq.~\re{charge} in 
the Appendix~\ref{app:Charges} and the corresponding definitions in 
Ref.~\cite{Adkins:1983hy}).} 
The iso\-scalar electric radius is more strongly  affected 
by the iso\-spin asymmetric environment
than the isovector one  because of  the presence of the 
isospin breaking term 
$\Delta^*$ (see appendix~\ref{app:Charges}). Consequently,
in proton-rich matter the ratio  
$\langle r^2\rangle^{*1/2}_{\rm E,I=0}/\langle r^2\rangle^{1/2}_{\rm E,I=0}$
is more enhanced than in neutron-rich matter.

\begin{figure}
\epsfysize=5.8cm
\epsffile{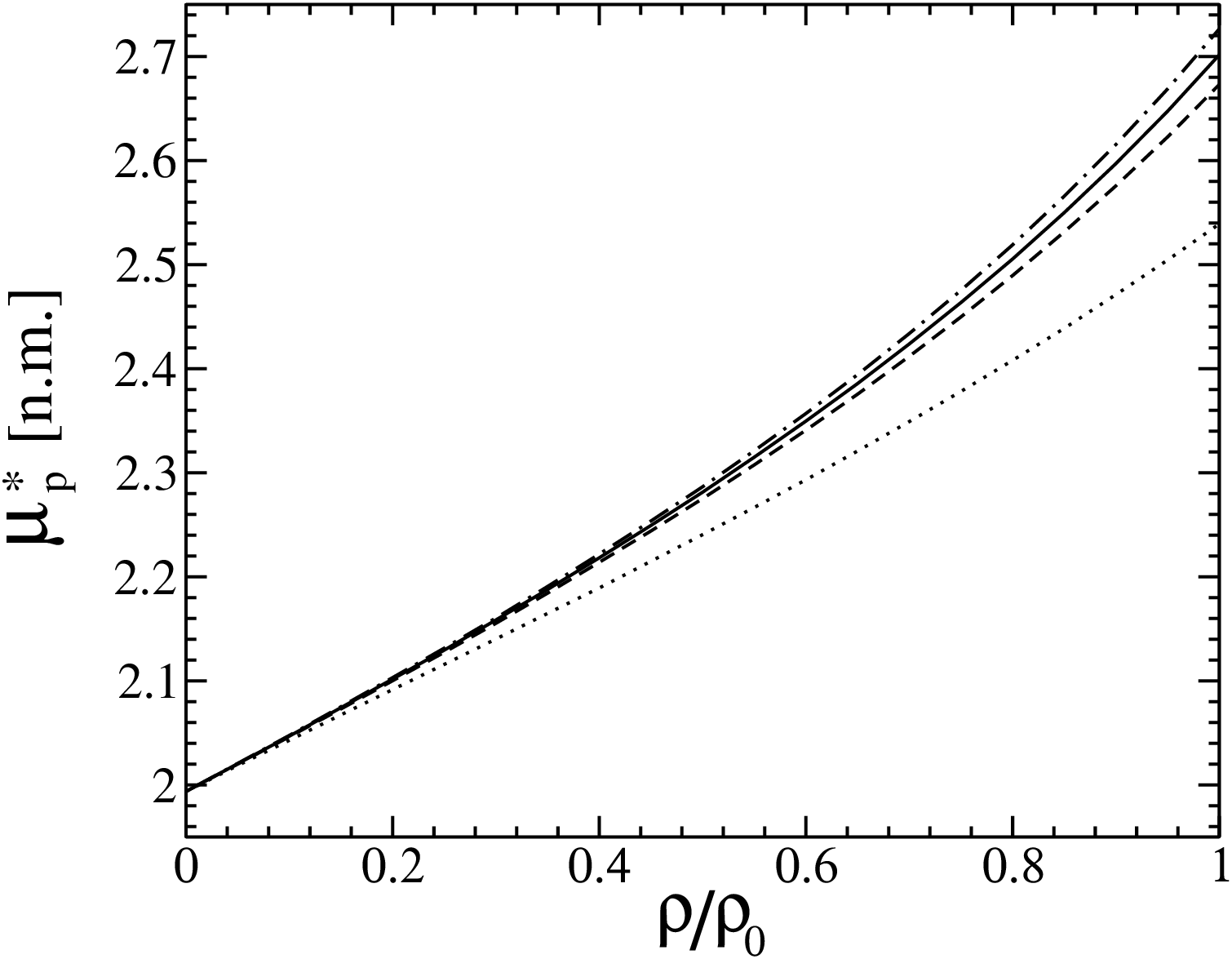}
\vspace{0.2cm}      
\caption{Density dependence of the proton magnetic moment.
The ordinate represents $\mu_{\rm p}^{*}$ in nuclear Bohr
magnetons (n.m.). 
The other definitions are as in Fig.~\ref{fig-strong}.}
\label{fig-muP}
\end{figure}

\begin{figure}
\epsfysize=5.8cm
\epsffile{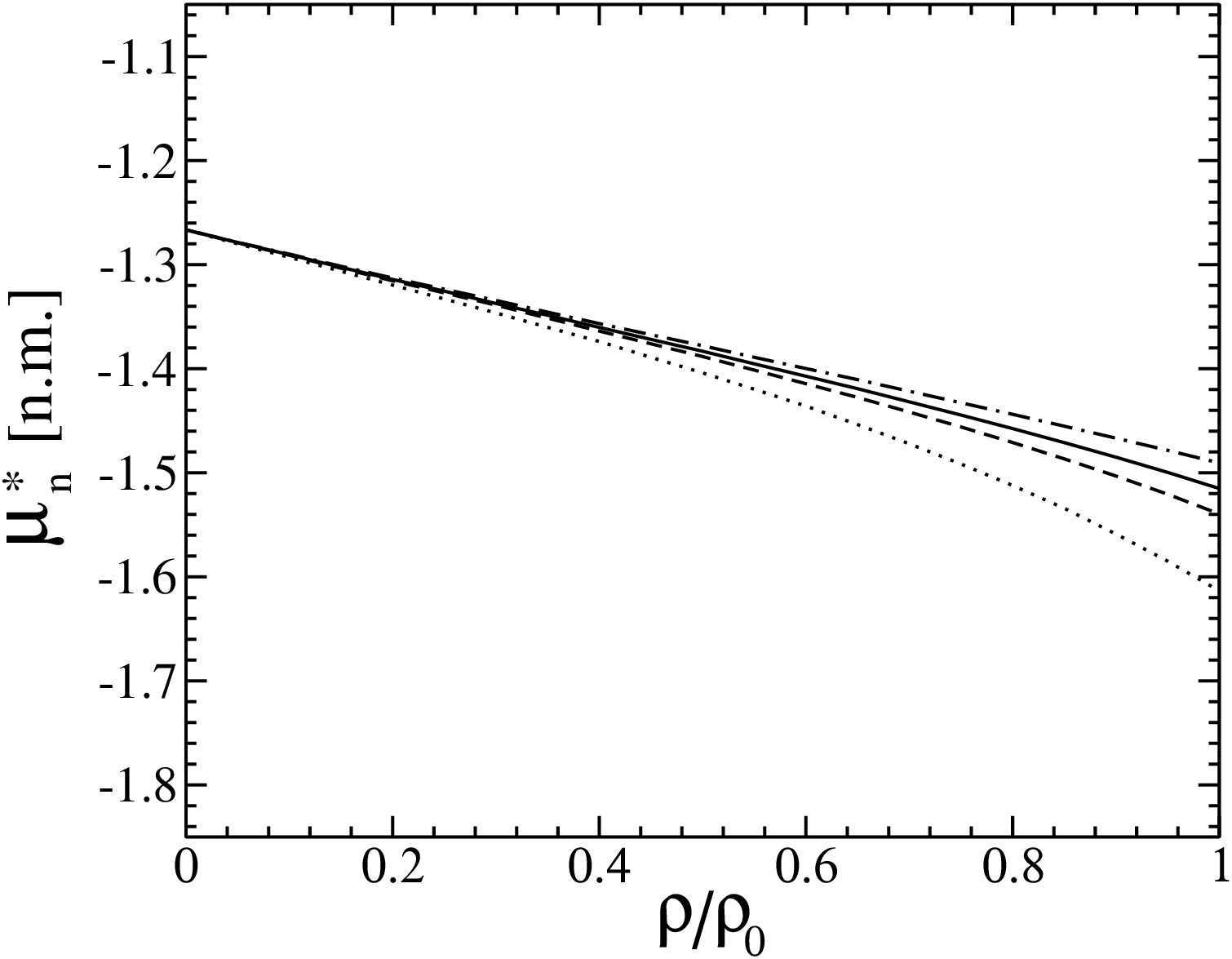}
\vspace{0.2cm}      
\caption{Density dependence of the neutron magnetic moment.
The ordinate represents  $\mu_{\rm n}^{*}$ in nuclear Bohr
magnetons (n.m.). 
The other definitions are as in Fig.~\ref{fig-strong}.}
\label{fig-muN}
\end{figure}

The density-dependence of the magnetic 
moments of the in-medium proton and the neutron is presented
in Fig.~\ref{fig-muP} and Fig.~\ref{fig-muN}, respectively. 
The influence of the
isospin asymmetry of the surrounding environment  on the in-medium magnetic
moments is comparatively weak in this case. Relative to the result in 
isospin-symmetric
matter both, the proton and neutron magnetic moments are 
decreased in neutron-rich matter and increased in proton-rich matter.

Let us conclude by remarking  
that within the present approach
the change of $\Delta m_{\rm np}^{*}$ 
is completely dictated by its strong part
when the isospin-asymmetry parameter is sizable
(compare Figs.~\ref{fig-strong}, \ref{fig-EM} and \ref{fig-total}).

\section{Summary and outlook}
\label{sect:summary}

We have proposed an effective Lagrangian which incorporates 
energy-dependent optical potentials for the $s$- and $p$-waves 
and which takes into account
the influence of the isospin-asymmetry of the environment 
onto the Skyrme-solitons of the model. 
As a result the neutron-proton mass splitting in 
asymmetric nuclear matter is predicted to vary strongly 
 relative to its free space value.  
The predictions obtained in the present work 
are in a qualitative agreement with the ones obtained
within relativistic hadron field theory
\cite{Li01,Gr0,Ho0}
and nonrelativistic variational calculations~\cite{Ch97}.
Quantitatively, however, the changes of  $\Delta m_{\rm np}^*$
are small in comparison 
to the results of those works.
Our approach shows that $\Delta m_{\rm np}^*$
in nuclear matter with sizable asymmetry
is mainly dictated by its strong part. 
In the case of more complicated calculations involving
finite nuclei  this may serve as a justification 
to evaluate only the strong part 
of the mass difference within the present approach.

Consequently, 
the next step in our future   
studies will be the estimate of $\Delta m_{\rm np}^*$
in finite  (particularly in mirror) nuclei. Here
additional effects are expected because the 
$p$-wave contribution proportional to 
$\Delta\chi_p\sim \vec\nabla^2\delta\rho$ 
in the Lagrangian~\re{lmedend} and the deformation effects discussed
in Refs.~\cite{Yakhshiev:2001ht,Yakhshiev:2002sr} 
become relevant.

\begin{acknowledgement}
The work of U.T.Y.\ was supported by the
Alexander von Humboldt Foundation.
Partial financial support from the EU Integrated Infrastructure
Initiative Hadron Physics Project (contract number RII3-CT-2004-506078),
by the DFG (TR 16, ``Subnuclear Structure of Matter'') and by BMBF
(research grant 06BN411) is gratefully acknowledged.
\end{acknowledgement}

\begin{appendix}

\section{Two-stage method and classical solutions}
\label{app:ClassSol}
\def\theequation{\Alph{section}.\arabic{equation}}
\setcounter{equation}{0}

In this appendix we review a new interpretation for the inclusion of the 
stationary $a^*$ rotations
that is
different from the one of Ref.~\cite{Rathske:1988qt} or the one
of Ref.~\cite{Meissner:2006id}, but which leads -- under the same input -- 
to the same results as in those references.
This new strategy is more convenient for our analytic calculations.

First of all we remark that stationary $a^*$ rotations  
essentially have to be introduced because of the
explicit isospin breaking in the mesonic sector, {\it i.e.} 
when ${\cal M}_-\ne 0$.  If
${\cal M}_-=0$ then $a^*=0$. Of course, the condition~\re{constr} 
satisfies this
requirement. Secondly, at the classical level, at which the soliton solution
is determined, this mesonic mass splitting can be ignored -- at the 
quantum level, however, this can not be the case 
since the symmetry breaking effect is enhanced 
by the coupling to the collective zero modes. Note the quadratic 
dependence on
the small parameter  ${\cal M}_-$ at the classical level, 
whereas at the quantum
level the dependence is linear; compare Eq.~\re{constr} with Eq.~\re{Ham}.

Let us, for the moment, 
put ${\cal L}_{\rm as}^*=0$ in the Lagrangian \re{lmedbeg}.
Then one can construct the classical 
`hedgehog' solution(s) from the static Lagrangian  
${\cal L}^{*\rm (static)}_{\rm sym}$.
In the work \cite{Meissner:2006id} these solutions are referred to as the
solutions of the non-perturbed (NP) system, determined by
\begin{equation}
\dsf{\delta M_{\rm NP}^*[F]}{\delta F}=0\,,
\label{NPsol}
\end{equation}
where $F(r)$ is the usual chiral profile function of the hedgehog 
ansatz and where 
$M_{\rm NP}^*=-\int {\cal L}^{*\rm (static)}_{\rm sym}\,{\rm  d}^3 r$ 
(compare with Eq.~\re{lag1}) 
is the in-medium mass of the soliton when it is not perturbed by any isospin
breaking.
The Lagrangian ${\cal L}^{*\rm (static)}_{\rm sym}$ is invariant under 
time-independent iso\-spin rotations
\begin{equation}
U\rightarrow AUA^\dagger,\quad A=\exp\{i\vec\tau\cdot{\vec\varphi}^{I}/2\}\,,
\label{A-rot}
\end{equation}
which can be treated as the usual zero modes of the model. The resurrection
of the time dependence ${\vec\varphi}^{I}\rightarrow{\vec\varphi}^{I}(t)$ in
Eq.~\re{A-rot}
leads to the (spatially integrated) Lagrangian
\[
\int {\cal L}^*_{\rm sym}\, {\rm d}^3 {r}=-M_{\rm NP}^*
+\dsf{\left(\dot{{\vec\varphi}}^{I}(t)\right)^2}{2}\Lambda^*\,,
\]
where $\Lambda^*$ is the in-medium moment-of-inertia.

Now we plug  the isospin breaking ${\cal M}_-\ne 0$ 
back into the mesonic sector, such that 
the corresponding  Lagrangian reads
\begin{equation}
\int{\rm d}^3 {r}\,\Big\{ {\cal L}^*_{\rm sym}+\Delta{\cal L}_{\rm mes}
\Big\}\no\\
=-M_{\rm NP}^*+\dsf{(\dot{\vec\varphi}^{I})^{2}}{2}\Lambda^*
-{\cal M}_-^2\Lambda_-\,.
\label{Lplug}
\end{equation}
Note that the integrated Lagrangian \re{Lplug}, even 
in the presence ${\cal M}_-\ne 0$, 
is still invariant under 
additional time-independent isospin rotations, {\it e.g.}
\begin{eqnarray}
A(t)
\to \tilde A(t)
&=&\exp\{i\vec\tau\cdot[{\vec \varphi}^{I}(t)+{\vec\varphi}^{II}]/2\}\no\\
&\equiv& \exp\{i\vec\tau\cdot {\vec \varphi}(t)/2\}\,. \label{Atilde}
\end{eqnarray}
In fact, it is a property of the hedgehog ansatz that 
these time-independent 
isospin rotations can be compensated by time-independent 
spatial rotations. The latter 
average out when 
the angular integration is performed, although  ${\cal M}_-\ne 0$.
But in order to be consistent in the use of 
the {\em unchanged} classical solution $F(r)$ determined by Eq.~\re{NPsol}, 
one has to introduce -- at the classical level -- the requirement
\begin{equation}
\dsf{(\dot{\vec\varphi}^{I})^{2}}{2}\Lambda^*-{\cal M}_-^2\Lambda_- 
 = 0\,.
 \label{classconstr}
\end{equation}
One can see explicitly that classically not all the rotations
$\varphi_1^{I},\varphi_2^{I},\varphi_3^{I}$ of Eq.~\re{A-rot}
can be time-independent. In other
words, at least one of the angular velocities 
$\dot\varphi_1^{I},\dot\varphi_2^{I},\dot\varphi_3^{I}$ 
must not vanish, say,
the one with respect to the third axis in isotopic space (which will later be
the quantization axis). We will call the
corresponding (via Eq.~\re{classconstr}) constrained angular velocity $a^*$,
such that
the condition \re{classconstr} is nothing else than the
constraint~\re{constr}.
As mentioned above, after angular averaging, the system remains
invariant under time-independent zero modes
which we called 
$\varphi_1^{II},\varphi_2^{II},\varphi_3^{II}$.
Again these zero modes can be made time-dependent
\begin{equation}
\dot\varphi_1^{II}=\omega_1,\quad
\dot\varphi_2^{II}=\omega_2,\quad
\dot\varphi_3^{II}=\omega_3,
\label{qrot0}
\end{equation}
where $\omega_1,\omega_2,\omega_3$, in contrast to $a^*$, 
are unconstrained angular velocities. Transcribed to our starting point 
\re{Atilde},
we eventually arrive at
\begin{equation}
\dot\varphi_1=\omega_1,\quad
\dot\varphi_2=\omega_2,\,\quad
\dot\varphi_3=\omega_3+a^*.
\label{qrot}
\end{equation}
Inserting $\dot\varphi_1,\dot\varphi_2,\dot\varphi_3$
into Eq.~\re{Lplug}
one generates the Lagrangian~\re{lag1} with $\Delta^*=0$. 
Interpreting now $\phi_1=\varphi_1,\phi_2=\varphi_2,\phi_3=\varphi_3-a^*t$ 
as collective coordinates 
and  $\omega_1,\omega_2,\omega_3$ as their pertinent velocities,
the standard quantization procedure 
should be performed around the stationary point
$a^* t$ with the constant angular velocity $a^*$.

This way of handling the classical $a^*$ rotations
is equivalent to the procedure
performed in Refs.~\cite{Meissner:2006id,Rathske:1988qt}.
But there is no need   anymore for the explicit 
introduction of the matrix ${\cal T}(t)$ (see equation (15) 
in Ref.~\cite{Meissner:2006id}) and the interpretation of the 
corresponding matrix order 
ambiguities discussed in Ref.~\cite{Rathske:1988qt}.

Thus at the analytical level one performs 
the usual Skyrme model calculations, where Eq.~\re{qrot} is inserted
into the expression \re{Lplug} 
and where the constraint~\re{constr} is applied at
the classical level, such that  $L^\ast= -M_{\rm NP}^*$ is extremized under
the hedgehog ansatz, see Eq.~\re{NPsol}. The latter corresponds to 
classical equation of motion \re{classic-eq} 
for the iso-symmetric matter case with 
vanishing $\delta\rho$  and therefore,
according to the definition~\re{polOpbe}, 
vanishing  $\Delta\chi_{s}$ (and 
vanishing $\Delta^*$, see Eq.~\re{clEe}).

Finally, the inclusion of isospin-breaking effects due to the 
asymmetric environment
($\Delta{\cal L}^*_{\rm env}\neq 0$) leads to the Lagrangian
\begin{eqnarray}
\lefteqn{\int{\rm d}^3 {r}\,\Big\{ {\cal L}^*_{\rm sym}
+\Delta{\cal L}_{\rm mes}+\Delta{\cal L}^*_{\rm env}
\Big\}}\no\\
&=&-M_{\rm NP}^*+\dsf{(\dot{\vec\varphi})^{2}}{2}\Lambda^*
-{\cal M}_-^2\Lambda_- + \dot\varphi\Delta^*\,.
\label{Lplug2}
\end{eqnarray}
Note that the Lagrangian \re{Lplug2}, because of the angular averaging, 
remains  invariant under
time-independent isospin-rotations as its counterpart  \re{Lplug}.
The collective coordinates are therefore
$\phi_1=\varphi_1,\phi_2=\varphi_2,\phi_3=\varphi_3-a^*t$ as before
and Eq.~\re{qrot} is still valid. When the latter
is inserted, Eq.~\re{Lplug2} transforms to the
Lagrangian~\re{lag1} which is the starting point for the quantization in
this work.
But in order to be consistent with the use of a solution of hedgehog type,
one still has to satisfy two requirements: 
(i) to apply the constraint \re{constr} in order
to remove $-{\cal M}_-^2\Lambda_-$ and 
$\Lambda^*{(\dot{\vec\varphi})^{2}}/2$ from \re{Lplug2},
and (ii) to extremize the remainder $-M_{\rm NP}^*+{\dot\varphi}\Delta^*$
(compare with Eq.~\re{lclassic})
under the hedgehog ansatz. Here $\dot\varphi(t)$ still corresponds to 
$\dot\varphi^{I}(t)=a^*$,
{\it i.e.} $\varphi^{II}$ is still assumed to be time-independent.
Consequently,  instead of Eq.~\re{NPsol} 
it is Eq.~\re{classic-eq} with  $\Delta\chi_{s} \neq 0$
that has to be solved (in practice
by iteration) in order to determine the profile function $F(r)$.

\section{Charges and magnetic moments}
\label{app:Charges}
\def\theequation{\Alph{section}.\arabic{equation}}
\setcounter{equation}{0}

In calculating the third component of the isospin current $V_0^{(3)}$,
one finds 
\begin{equation}
\int{\rm d}^3{r}V_0^{(3)}
=(\omega_3+a^*)\Lambda^*+\Delta^*\equiv T_3\,.
\end{equation}
Consequently, the isospin charge {\em density} -- 
modulo a factor $4\pi r^2$ -- 
is given as
\begin{equation}
\tilde T_3=(\omega_3+a^*)\tilde\Lambda^*+\tilde\Delta^*
=(T_3-\Delta^*)
\dsf{\tilde \Lambda^*}{\Lambda^*}+\tilde\Delta^*\,.
\end{equation}
Note that there are terms that are independent of 
the isospin $T_3$ on the r.h.s.
Since the charges of the nucleons are defined as 
\begin{equation}
Q=\dsf{B}{2}+T_3\equiv\int\limits\rho_{I=0}(r)\,{\rm d}^3r\pm
\int\limits\rho_{I=1}(r)\,{\rm d}^3r\,,\label{charge}
\end{equation}
the isoscalar and the isovector density distributions have here 
the following form
\begin{eqnarray}
4\pi r^2\rho_{I=0}(r)&=&\dsf{\tilde B}{2}-\dsf{\Delta^*}{\Lambda^*}
{\tilde\Lambda^*}+\tilde\Delta^*\,,\\
4\pi r^2\rho_{I=1}(r)&=&\dsf{\tilde\Lambda^*}{2\Lambda^*}\,.\\
&&\no
\end{eqnarray}
Analogous calculations for the magnetic moments lead to 
\begin{eqnarray}
\mu^{\left({\rm p}\atop {\rm n}\right)*}&=&\dsf{m_{\rm N}
(1+\Delta^*)}{6\Lambda^*}
\int\limits_0^\infty\tilde Br^2{\rm d}r\no\\
&&\mbox{}\pm\dsf{m_{\rm N}}{3}
\int\limits_0^\infty\left[\big(1-\chi_{\rm p}^0\big)\tilde\Lambda_-
+\tilde\Lambda_4^*\right]{\rm d}r\,.
\end{eqnarray}

\end{appendix}

\end{document}